\documentclass[prb,preprint]{revtex4}
\usepackage{epsfig}
\usepackage{amsmath}
\usepackage{graphicx}

\begin{document}

\newcommand{\bq}{{\bf q}}

\newlength{\figwidth}
\figwidth=0.8 \textwidth
\newcommand{\fg}[3]
{\begin{figure}[tb]
{\includegraphics[width=\figwidth]{#1}}
\caption{#2}\label{#3}\end{figure}}
\newcommand{\fgb}[3]
{\begin{figure}[b]
{\includegraphics[width=\figwidth]{#1}}
\caption{#2}\label{#3}\end{figure}}
\newcommand{\fgz}[3]
{\begin{figure}[here]
{\includegraphics[width=\figwidth]{#1}}
\caption{#2}\label{#3}\end{figure}}

\title{The Casimir effect from a Condensed Matter Perspective}

\author{L. P\'alov\'a}
\author{P. Chandra}
\author{P. Coleman}
\affiliation{Center for Materials Theory, Department of 
Physics and Astronomy, Rutgers University, 
Piscataway, New Jersey 08854}


\begin{abstract}

The Casimir effect, a key observable realization of vacuum
fluctuations, is usually taught in graduate courses on quantum field
theory. The growing importance of Casimir forces in
microelectromechanical systems motivates this subject as a topic
for graduate many-body physics courses. To this end, we revisit the
Casimir effect using methods common in condensed matter physics. We
recover previously derived results and explore the
implications of the analogies implicit in this treatment.
\end{abstract}

\maketitle

\section{Introduction}
The Casimir effect results from the interplay of zero-point fluctuations
and boundary conditions, and leads to the attraction between two 
parallel conducting plates 
in a vacuum.\cite{Casimir48,Lambrecht02,Lamoreaux07} 
It was one of the first predicted, observable consequences 
of vacuum fluctuations. Traditionally it has been taught in graduate
courses on quantum field theory.\cite{Itzykson80}
Because the Casimir force scales inversely proportionally to 
the fourth power of the
plate separation $a$, it is only measurable when $a$ is 
quite small (micron regime). Recently, the 
Casimir phenomenon has assumed a new importance in the design of nanoscale 
devices.\cite{Lamoreaux97,Mohideen98,Chan01,Lisanti05,Obrecht07,Munday09} 
Generalizations
of the Casimir effect to critical systems with external constraints 
continue to be 
fruitful.\cite{Fisher78,Krech94,Kardar99,Danchev00,Gambassi08,Palova08}
It therefore is timely to include it 
in graduate courses on condensed matter theory.

In this paper
we revisit the Casimir effect, recovering previously derived 
results \cite{Casimir48,Belifante87,Dowling89,Elizalde91,Revzen97,Lamoreaux99} 
with an approach 
used to calculate observable properties in finite-temperature solids, and 
can therefore be naturally included in the many-body 
curriculum,\cite{Mahan07} possibly serving as a simple pedagogical
example of these techniques.

\section{The Casimir Coefficient}
The Casimir effect results from the effect of boundary conditions on
the zero-point fluctuation 
modes of the electromagnetic field.
We will consider the simplest case of 
two parallel conducting
plates. The energy, 
$\Delta {\cal E}$, is the finite difference between 
the zero-point energies with and without 
the plates,\cite{Casimir48,Belifante87,Dowling89,Elizalde91,Revzen97} and
the force is then the spatial derivative of $\Delta {\cal E}$.
The component of the electric field parallel to the conducting plates
must vanish. There are two sets of modes that satisfy this
condition: the transverse electric (TE) and transverse magnetic (TM) modes
where the electric or magnetic field are respectively parallel to the
plates.\cite{Jackson98}
The electric field for the transverse electric
field modes is given by
\begin{equation}
\vec{E}^{{\rm TE}} (\vec{x},z) = 
\sum_{\vec{q}_{\perp },n>0}
{E}_{\vec{q}n} (\hat z\times \hat q_{\perp })
e^{i\vec{q}_{\perp}\cdot\vec{x}} \sin
\left(\frac{n\pi}{a}z \right), \qquad (n>0)
\end{equation}
where $\vec{x}$ and $z$ are the co-ordinates parallel and perpendicular to the plates respectively, $n$ is an integer, and 
${E}_{\vec{q}n}$ is the Fourier amplitude of the fields.
There is no $n=0$ TE mode. The corresponding magnetic field is
calculated using Faraday's equations $\vec{\nabla}\times \vec{E} =
- \partial \vec{B}/\partial t$, or $\vec{B} = \dfrac{1}{i\omega}\vec{ \nabla
}\times \vec{E}$.
The magnetic field for the TM field modes
is given by 
\begin{equation}
\vec{B}^{{\rm TM}} (\vec{x},z) = 
\sum_{\vec{q}_{\perp },n>0}
{B}_{\vec{q}n} (\hat z\times \hat q_{\perp })
e^{i\vec{q}_{\perp}\cdot\vec{x}} \cos
\left(\frac{n\pi}{a}z \right), \qquad (n\geq 0).
\end{equation}
We note that there is one extra $n=0$ TM mode. The corresponding electric fields are computed from Maxwell's
displacement current equation $\vec{\nabla }\times \vec{B} =
\dfrac{1}{c^{2}}\dfrac{\partial \vec{E}}{\partial t}$ or $\vec{E}= -
\dfrac{c^{2}}{i\omega} (\vec{\nabla }\times \vec{B})$. 
The
Fourier modes of these fluctuations thus involve a discrete set of wavevectors, 
\begin{equation}\label{}
\vec{q}_{n} = (\vec{q}_{\perp }, q_{zn}), 
\end{equation}
where $q_{zn} = n\pi/a$ and $n$ is an integer, 
leading to a discrete
set of normal mode frequencies 
$\omega_{\vec{q}_{\perp}n} = c \sqrt{\vec{q}_{\perp}^2 + q^2_{zn}}$ (see Fig.~\ref{fig1}),
where $c$ is the speed of light.

\fg{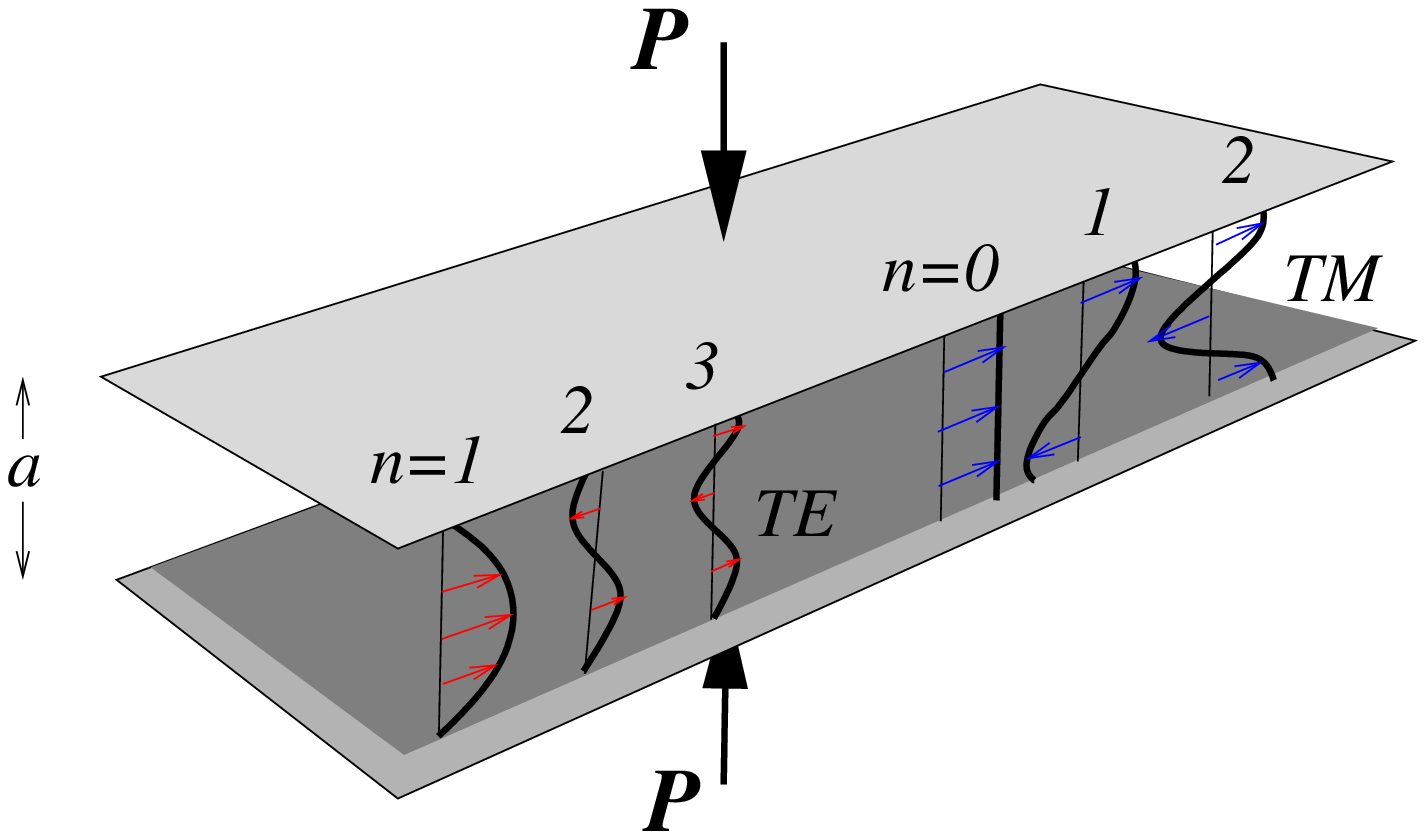}{Schematic of the Casimir effect indicating that
the normal 
modes of the electromagnetic field between two conducting plates occupy a
discrete set of wavevectors. In the transverse
electric (TE) modes the electric field lies parallel to the plates
and $n\ge 1$. In the TM modes the magnetic
field lies parallel to the plates and $n\geq 0$.
The modification of 
the frequencies of the 
zero-point fluctuations by boundary conditions changes the 
energy of the system, creating a pressure
on the plates.}{fig1}

The zero-point energy of the fields inside the plates is given by 
\begin{equation}
{\cal E}_C = \sum_{\vec{q}_{\perp}} \frac{\hbar
\omega_{\vec{q}_{\perp},0}}{2} +
2 \sum_{\vec{q}_{\perp},n>0} \frac{\hbar
\omega_{\vec{q}_{\perp},n}}{2},
\end{equation}
where the first term is the zero point energy of the $n=0$ TM mode,
and the second term counts the zero point energy of the TM and TE modes
with $n>0$. 
We may rewrite these two terms as a sum over all $n$, both
positive and negative, as follows
\begin{equation}
\label{Ec}
{\cal E}_C = \frac{\hbar c}{2} 
\sum_{n = -\infty }^{\infty }\sum_{\vec q_{\perp }}
\sqrt{q_{\perp}^2 + q_{zn}^2}.
\end{equation}
In the continuum limit we let $\sum_{\vec{q}_{\perp }}\rightarrow
A\!\int\! \dfrac{d^{2}q_{\perp }}{(2\pi)^{2}}$, where $A$ is the area of the
plates, to obtain
\begin{equation}
\label{Ec}
{\cal E}_C = 
A \frac{\hbar c}{2}
\sum_{n = -\infty }^{\infty }\int\!\frac{d^2 q_{\perp}}{(2 \pi)^2} \sqrt{q_{\perp}^2 + q_{zn}^2}.
\end{equation}
The quantity ${\cal E}_C/A$ determined from Eq.~(\ref{Ec}) 
is dimensionally of the form $[{\cal E}_{C}/A]=\hbar c [L^{-3}]$.
Because $a$ is the only length scale in the
system, it follows that the change in the zero-point energy must
have the form
\begin{equation}
\frac{\Delta {\cal E}_c}{A} = {\cal K} \frac{\hbar c}{a^3}.
\label{Escaling}
\end{equation}
The fact that this Casimir energy is sensitive to arbitrary interplate 
separation, $a$, is a direct consequence
of the gaplessness, and thus the scale-free nature of the photon field.

The traditional calculation of ${\cal K}$ in the Casimir 
energy Eq.~(\ref{Escaling}) is performed using a regularization procedure 
enforced by a zeta 
function.\cite{Casimir48,Belifante87,Elizalde91}
In this paper we present an alternative derivation in which we
calculate the necessary sums by exploiting the structure of
the Bose function and the residue theorem of complex analysis.
This approach is central to the
Matsubara formalism\cite{Mahan07} used to study many-body systems at
finite temperature. 
The calculation of ${\cal K}$ by contour integration
has been discussed,\cite{Dowling89,Revzen97} 
and we will
adapt this treatment as an example
of the Matsubara method\cite{Mahan07} in a graduate course
in many-body physics.
Therefore we take a brief diversion to 
describe the technique in general before applying it specifically to
the calculation of the Casimir coefficient.

\section{The Matsubara Approach}

We begin by noting that the Bose function
\begin{equation}
\label{Bose}
n_B(z) = \frac{1}{e^{\hbar z/k_BT} - 1}
\end{equation}
has poles on the imaginary axis (see Fig.~2) at
$
z = i\nu_n$, where $\nu_n 
= n 2 \pi k_B T/\hbar$,
because
\begin{equation}
e^{i \hbar \nu_n/k_B T} = e^{2 \pi n i} = 1.
\end{equation}
Next we take 
\begin{equation}
\label{zdelta}
z = i\nu_n + \delta,
\end{equation}
where $\delta$ is small so that $z$ is slightly off the imaginary axis so that
\begin{equation}
\label{ndelta}
n_B(i\nu_z+\delta) = \frac{1}{e^{\hbar \delta/k_BT} - 1} \approx \frac{k_BT}{\hbar \delta},
\end{equation}
from which we see that $k_BT$ is the residue at each of the poles
$z = i\nu_n$ of $n_B(z)$, so that 
\begin{equation}
\label{nBpoles}
n_B(z) = \sum_n \frac{k_B T}{\hbar (z - i\nu_n)}.
\end{equation}

If we have a function, $F(z)$, that does not have poles on the imaginary
axis, we can use the residue theorem and Eq.~(\ref{nBpoles}) to 
write
\begin{equation}
\label{residue}
\oint_{\cal C} dz F(z) n_B(z) 
= 2 \pi i \sum_n \frac{k_B T}{\hbar} F(i\nu_n),
\end{equation}
where ${\cal C}$ is a contour that encircles the imaginary axis in a
clockwise
sense, 
as shown
in Fig.~\ref{fig2}. Equation~(\ref{residue})
can be rearranged to read
\begin{equation}
\label{Fsumint}
\sum_n F(i\nu_n)
= \frac{\hbar}{k_B T}
\oint_{\cal C} \frac{dz}{2\pi i} F(z) n_B(z),
\end{equation}
which is a key result in the Matsubara approach used to 
evaluate sums that emerge in the study
of many-body systems at finite temperatures.\cite{Mahan07}

\figwidth=4.0in
\fg{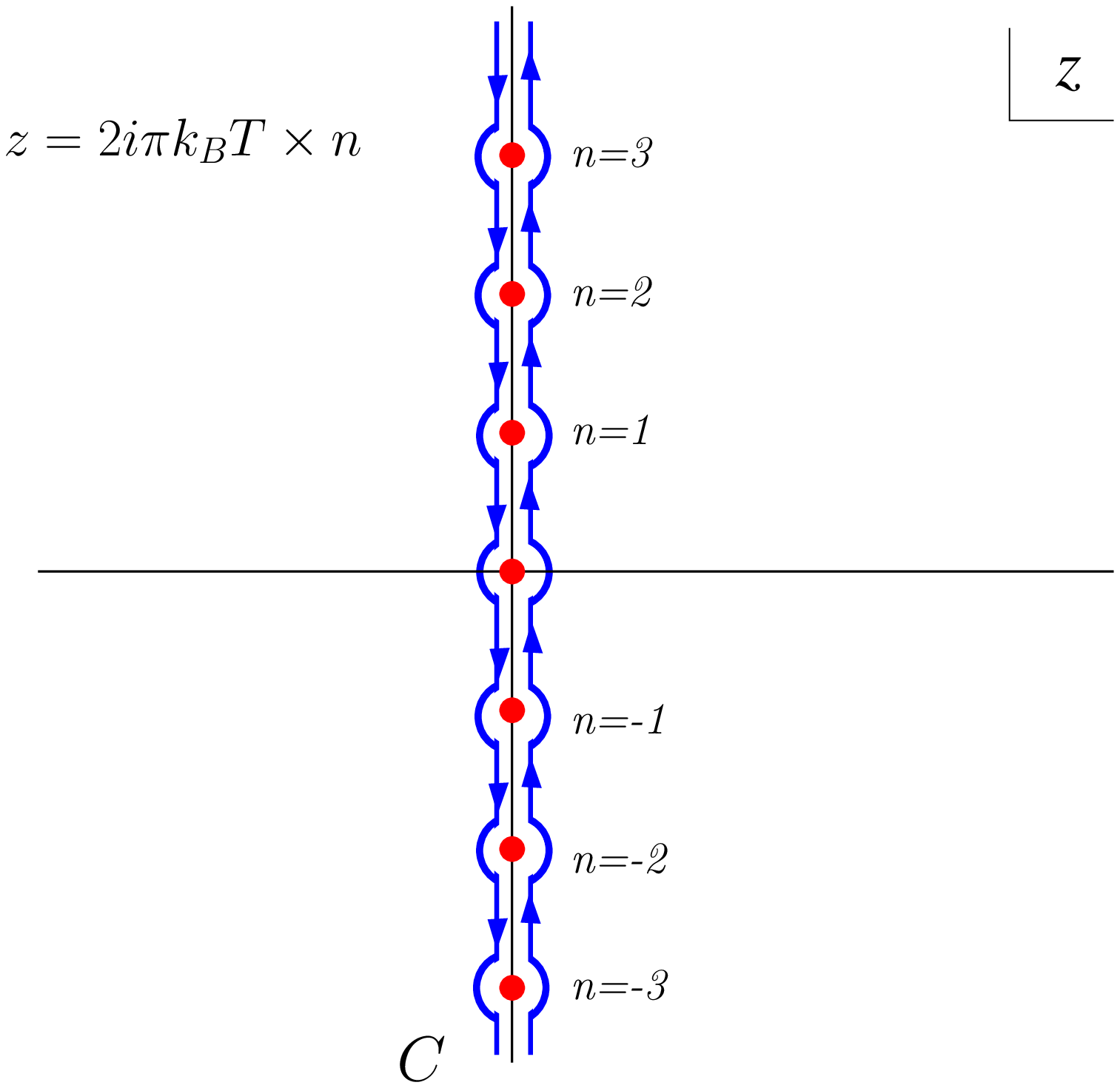}{Contour integration path $\cal C$ used to
sum over the Matsubara frequencies.}{fig2}

We will now apply Eq.~(\ref{Fsumint}) to the specific 
case of the Casimir coefficient.
To do so, we must identify the summations
that we need to calculate.
We begin with the zero-point energy per unit area (\ref{Ec}) 
\begin{equation} 
\frac{{\cal E}_C }{A} 
=
\frac{\hbar c}{2}
\sum_{n}
\!\int \frac{d^2 q_{\perp}}{(2 \pi)^2} \sqrt{q^2_\perp + q_{zn}^{2}}.
\label{Ec2}
\end{equation}
We are interested only in the {\it change} in the zero point
energy as a result of the plates. In the limit of 
infinite plate separation ($a\rightarrow
\infty $) the discrete interval in $q_{zn}$, $\Delta q_{zn} = \pi/a$,
becomes infinitesimal, and the
sum over $n$ in Eq.~(\ref{Ec2}) can be replaced by an integral
$\sum_{n}= \sum_{n}\dfrac{\Delta q}{\pi/a} = \dfrac{a}{\pi}\!\int dq_z$.
Therefore 
the change in the zero-point energy per unit area due to the
presence of the plates is given by
\begin{eqnarray}\label{Iqa}
\frac{\Delta {\cal E}_C }{A} 
= \frac{{\cal E}_C }{A} - \left. \frac{{\cal E}_C }{A}
\right\vert_{a\rightarrow \infty}
= {\hbar ca}\!
\int \frac{d^2 q_{\perp}}{(2 \pi)^2} \ I (q_{\perp},a),
\end{eqnarray}
where 
\begin{eqnarray}\label{Ec3}
I (q_{\perp},a)
&=&\frac{1}{2a}\sum_{n}\sqrt{q_{\perp}^{2}+q_{zn}^{2}}
-\!\int \frac{dq_{z}}{2\pi}
\sqrt{q_{\perp}^{2}+q_{z}^{2}}.
\end{eqnarray}
By making this subtraction, we remove the ultraviolet divergences in
the zero-point energy.
By using the Matsubara method, we can reexpress 
the sum in 
Eq.~(\ref{Ec3}) as 
\begin{equation}
\label{sum}
\frac{1}{2ac}\sum_{n }\sqrt{(c q_{\perp})^{2}+ (c q_{zn})^{2}}
= \frac{1}{\hbar c^{2}\beta_{C}}\sum_{n} F(i\nu_n),
\end{equation}
where 
\begin{equation}
\label{function}
F (z)= \sqrt{c^2 q_{\perp}^2 - z^2} .
\end{equation}
We associate the discrete wavevectors, 
$q_{zn}$, with a ``Matsubara frequency'' $c q_{zn}\equiv \nu_n$.
Then
\begin{equation}
c q_{zn}= c n \frac{\pi}{a} \equiv 
n \left(\frac{2 \pi k_B T_{C} } {\hbar}\right),
\end{equation}
where the effective Casimir temperature is given by 
\begin{equation}
k_B T_{C}=\frac{\hbar c}{2a},
\label{effcass}
\end{equation}
so that 
\begin{equation}\label{betac}
\beta_{C}= \frac{1}{k_B T_{C}}\equiv \frac{2a}{\hbar c}.
\end{equation}
We note that $T_{C}$ scales
inversely with the plate separation
($T_C \sim 1/2a$).

Following the Matsubara approach, the sum in Eq.~(\ref{sum}) can now 
be rewritten as a contour 
integral\cite{Mahan07} 
around the poles at
$z=i\nu_n$
of the Bose function 
$n_{B}(z,\beta_C)$ yielding 
\begin{equation}
\frac{1}{\hbar \beta_{C}}\sum_{n} F(i \nu_n) =\!\int_C \frac{dz}{2\pi i} F(z)
n_B(z,\beta_{C}).
\label{Poisson}
\end{equation}
\figwidth=6.0in
\fg{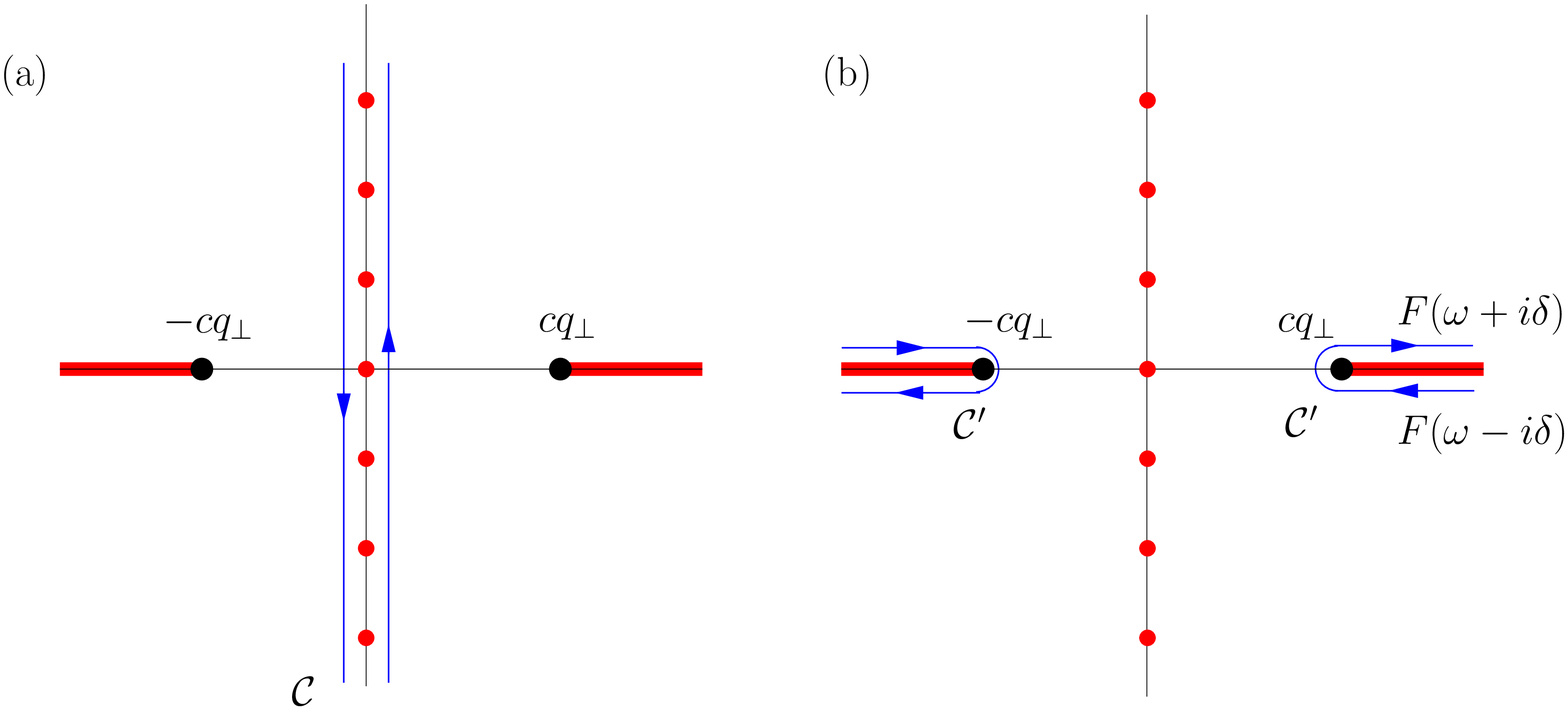}{(a) Contour integration path $\cal C$ used to
calculate $\dfrac{1}{\hbar \beta_{C}}\sum_n F (i\nu_n)$ in
Eq.~(\ref{Poisson}), where $F (z)= \sqrt{c^{2}q_{\perp }^{2}-z^{2}}$, 
showing branch cuts in $F (z)$ at $z= \pm c
q_{\perp }$. (b) Distortion of the contour into contour $C'$ that wraps around the
branch cuts of $F (z)$. The integrand of the integral along the
branch cuts
is the difference $F (\omega+i\delta) -F (\omega-i\delta)$ between
the value of $F (z)$ above and below the branch cut.
}{fig3}
The second term in $I (q_{\perp },a)$, Eq.~(\ref{Ec3}), corresponds to the
$a\rightarrow \infty $, or $\beta_{C}\rightarrow \infty $ limit of the
first term, and thus we may write
\begin{equation}\label{reg}
I (q_{\perp },a) = \lim_{\beta'\rightarrow \infty }
\left\{ \frac{1}{c^{2}}\!\int_C \frac{dz}{2\pi i} F(z)
\left[ n_B(z,\beta_C)-
n_{B} (z,\beta ')\right]\right\}.
\end{equation}

The subtraction of the $\beta_C\rightarrow \infty $ 
limit of the integrand in Eq.~(\ref{reg})
regulates the overall integral at large $z$, guaranteeing that the integrand
around a contour at infinity vanishes. This procedure permits us to
evaluate the integral by distorting the 
contour around the branch cuts in $F (z)$ that extend from $z = \pm
c q_{\perp }$ to infinity, as shown in Fig.~(\ref{fig3}). We then obtain
\begin{equation}\label{toinfinityandbeyond}
I (q_{\perp},a) =\frac{1}{c^{2}}\left(\int_{-\infty }^{-cq_{\perp }} +
\int_{cq_{\perp }}^{\infty }\right) 
\frac{d\omega}{2 \pi i} [F (\omega+i\delta)-F (\omega-i\delta)]
\left[n_{B} (\omega,\beta_{C}) -\{\beta_{C}\rightarrow \infty \} \right].
\end{equation}
To evaluate the branch cut, we note for 
$F (\omega \pm i \delta)= \sqrt{(cq_{\perp })^{2}- (\omega\pm i \delta
)^{2}}$,
\begin{equation}\label{}
F (\omega +i \delta)-F (\omega-i\delta) = -2 i \sqrt{\omega^{2}- (cq_{\perp})^{2}}\ {\rm sgn} (\omega),
\end{equation}
is an odd function of $\omega$, which permits us 
to replace $n_{B} (\omega)$
by its odd part $n_{B} (\omega)+\dfrac{1}{2}$ 
to obtain
\begin{equation}\label{toinfinityandbeyond}
I (q_{\perp},a) =\frac{1}{c^{2}}\left(\int_{-\infty }^{-cq_{\perp }} +
\!\int_{cq_{\perp }}^{\infty }\right) 
\frac{d\omega}{ 2\pi i } [F (\omega+i\delta)-F (\omega-i\delta)]
\left[\{ n_{B} (\omega,\beta_{C})+\frac{1}{2}\} -\{\beta_{C}\rightarrow \infty \} \right].
\end{equation}
Because the integrand is an even function of $\omega$, we can
replace this integral by twice the integral over positive $\omega$ to obtain
\begin{eqnarray}\label{toinfinity}
I (q_{\perp},a) &=& - \frac{2}{c^{2}}\!\int_{c q_{\perp}}^{\infty }
\frac{d\omega}{\pi}\sqrt{\omega^{2}-c^2 q_{\perp}^{2}} \, 
\left[\left( n_{B} (\omega,\beta_{C})+ \frac{1}{2} \right) -
\left\{\beta_C \rightarrow \infty \right\} \right]\cr
&=& - \frac{2}{c^{2}}\!\int_{c q_{\perp}}^{\infty }
\frac{d\omega}{\pi}\sqrt{\omega^{2}-c^2 q_{\perp}^{2}} \, n_{B} (\omega,\beta_{C}).
\end{eqnarray}

The change in zero point energy is then given by 
\begin{equation}
\frac{\Delta {\cal E}_C}{A} = 
- 2 \hbar^2 \beta_C\!\int_{\omega>q_{\perp}} 
\frac{d^{2}q_{\perp} d\omega}{(2\pi)^{3}}n_B(\omega,\beta_C)\sqrt{\omega^{2}-c^2 q_{\perp}^{2}},
\end{equation}
where we have made the substitution $2a/c=\hbar \beta_{C}$.
By carrying out the integral over $q_{\perp}$, we obtain
\begin{equation}
\frac{\Delta {\cal E}_C}{A} = - 
\frac{\hbar^2 \beta_C}{6 \pi^{2} c^2} 
\!\int d\omega \omega^{3}
n_B(\omega,\beta_C).
\label{bbody}
\end{equation}

Rescaling the
integral in Eq.~(\ref{bbody}) 
by changing variables to $x = \hbar\omega/k_B T$
and replacing $\beta_C = 2a/\hbar c$, we obtain 
\begin{equation}\label{coefficient}
\frac{\Delta {\cal E}_C}{A} = - 
\frac{1}{6 \pi^2 \hbar^2 \beta_C^3 c^2} \overbrace {\int dx \frac{x^{3}}{e^{x}-1}}^{\frac{\pi^{4}}{15}}
=- \frac{\pi^2}{720} \frac{\hbar c}{a^3}
\end{equation}
in numerical agreement with previous 
derivations.\cite{Casimir48,Belifante87,Dowling89,
Elizalde91,Revzen97}
The associated force per unit area is then
\begin{equation}
\label{force}
\frac{F}{A} = \frac{d \Delta {\cal E}_c}{d a} 
= \frac{\pi^2}{240} \frac{\hbar c}{a^4} = 1.3 \times 10^{-3}
\frac{1}{(a/ \mu {\rm m})^4} 
N/m^2
\end{equation}
indicating that measurements of the Casimir force 
must be performed at plate separations at or below
the micrometer length scale.
\cite{Lamoreaux97,Mohideen98,Chan01,Lisanti05,Obrecht07,Munday09} When the
two conducting plates are parallel, the force is attractive, but
it can be repulsive in other situations.\cite{Boyer74,Hushwater97}

\section{Implications}

We end with a discussion of the broader implications of our
approach. Here we use the Bose function
$n_{B} (\omega)$ to
impose a spatial boundary condition on quantum fluctuations. Key 
to this treatment is the interpretation of 
the discrete $q_{zn}$ vectors as a set of
Matsubara frequencies $cq_{zn}\equiv \nu_n$ so that we are effectively mapping 
the spatial frequencies of the Casimir effect 
to a discrete set of {\it temporal} frequencies in statistical
mechanics.
This procedure leads to an effective Casimir temperature Eq.~(\ref{effcass}), 
$k_{B}T_{C}=
\hbar c/2a$, 
where $T_{C}$ is linked to a 
spatial boundary
condition. If we now return to statistical mechanics, 
our treatment of the Casimir effect allows us to
illustrate a deep relation between finite
temperature and boundary conditions in time, not in space.
This relation arises from the fact that the
Boltzmann factor $e^{-\beta \hat H}$ 
corresponds
to the unitary time evolution operator of quantum mechanics
\begin{equation}
e^{-\beta \hat H}= e^{-\frac{i \hat H}{\hbar} [-i\hbar \beta]} = U [-i\hbar \beta]
\label{unitary}
\end{equation}
evaluated at the imaginary Planck time
$t= -i \hbar \beta = - i\hbar/k_B T$.
Thus statistical mechanics can 
be formulated as quantum mechanics in
imaginary time, where the temporal evolution occurs 
along the imaginary time axis with 
$t= -i \tau$, where $\tau \in [0,\hbar \beta ]$.

To give the reader a flavor for this link between finite temperature
and temporal boundary conditions, we 
consider a simple harmonic oscillator
described classically by the action
\begin{equation}
S_{\rm cl } =\!\int_{t_{1}}^{t_{2}}dt\ \left[\frac{m\dot\phi^{2}}{2}
-\frac{m\omega_{0}^{2}\phi^{2}}{2}
\right],
\end{equation}
where $\phi(t)$, 
$m$ and $\omega_{0}$ are its amplitude, mass, and angular frequency respectively.
The passage from classical to quantum mechanics is achieved
using Feynman's observation\cite{Feynman65} that the amplitude for the oscillator to 
follow a path 
$\phi (t)$ is $\exp[i S_{\rm cl}/\hbar]$.
If $| \phi_{1}\rangle$ is the eigenstate of displacement,
then the transition amplitude between 
$\vert \phi_{1}\rangle $ and $\vert \phi_{2}\rangle $ is the sum 
\begin{equation}
\langle \phi_{2}\vert U (t_{2}-t_{1})\vert \phi_{1}\rangle =
\sum_{\{\phi (t) \}} \exp \left[\frac{i
}{\hbar }S_{\rm cl}\right].
\end{equation}
over all paths 
that link $\phi_{1}$ and $\phi_{2}$.
We pass from real to imaginary time,
and identify $t\rightarrow -i\tau$ and 
the Euclidean action $S_E$ by
$\dfrac{i}{\hbar } S_{\rm cl} \rightarrow -\dfrac{1
}{\hbar }S_{E} $. By making the necessary replacements,
we find that 
the imaginary time evolution from $\vert \phi_1\rangle $ to $\vert
\phi_{2}\rangle $ is associated with the amplitude
\begin{equation}
\langle \phi_{2}\vert e^{-\beta \hat H}\vert \phi_{1}\rangle =
\sum_{\{\phi : \ \phi (0)=\phi_{1},\ \phi ( \hbar \beta) = \phi_{2} \}} \exp \left[-\frac{1
}{\hbar }S_{E}\right].
\end{equation}
To obtain the associated partition function, we must 
take the trace over this
matrix
\begin{subequations}
\begin{align}
Z &= {\rm Tr}[e^{-\beta \hat H}]=\!\int_{-\infty }^{\infty } d\phi
\langle \phi \vert e^{-\beta \hat H}\vert \phi \rangle = 
\!\int d\phi \sum_{\phi =\phi (0) = \phi (\hbar \beta) } \exp \left[-\frac{1
}{\hbar }S_{E}\right] \\
&= \sum_{\phi(\hbar\beta)=
\phi (0) } 
\exp \left ( -\frac{1
}{\hbar }S_{E}[\{\phi \}]\right),
\end{align}
\end{subequations}
where the sum is over all periodic paths that satisfy 
$\phi (0)= \phi (\hbar \beta)$. 
We see that a finite temperature is formally equivalent to
a periodic boundary condition in imaginary time. We next 
apply these ideas to our illustrative case. 

In the passage from real to imaginary time,
the quantum mechanical amplitude has been replaced by a ``probability''
$p[\{\phi \} ] \propto e^{-S_{E}[\phi ]/\hbar}$.
To calculate this function for our simple example, 
we decompose the displacement in terms of its normal modes
(Fig.~\ref{pedfig4}) by
Fourier transforming in imaginary time
\begin{equation}
\phi (\tau)= \frac{1}{\sqrt{\hbar \beta }}\sum_n \phi_{n}e^{-i \nu_n\tau },
\end{equation}
where $\nu_n = 2\pi n/\hbar \beta$ and $\phi_{n}^{*}= \phi_{-n}$ 
because $\phi(\tau)=\phi(\tau)^*$ is real.
We can then write the Euclidean action as a sum of contributions from
each normal mode
\begin{equation}
\frac{S_{E}[\{\phi \}]}{\hbar }= \frac{m}{2\hbar }\sum_{n=-\infty}^{\infty}
\vert \phi_{n}\vert^{2}\left[\omega_{0}^{2}+ \nu_n^2 \right],
\end{equation}
so that the probability amplitude $p[\{\phi \}]= Z^{-1} e^{-S_{E}/\hbar}$
factorizes into a 
product of Gaussian distribution functions for each normal mode:
\begin{equation}\label{l}
p[\{\phi \} ]\propto \exp \left[- \sum_n \frac{\vert \phi_{n}\vert^{2}}{2\sigma_{n}^{2}} \right],
\end{equation}
where
\begin{equation}
\sigma_{n}^{2}= \langle \vert \phi_{n}\vert^{2}\rangle = \frac{\hbar }{m (\omega_{0}^{2}+\nu_n^2)}
\end{equation}
is the temperature-independent variance of each normal mode.
\figwidth=7.0in
\fg{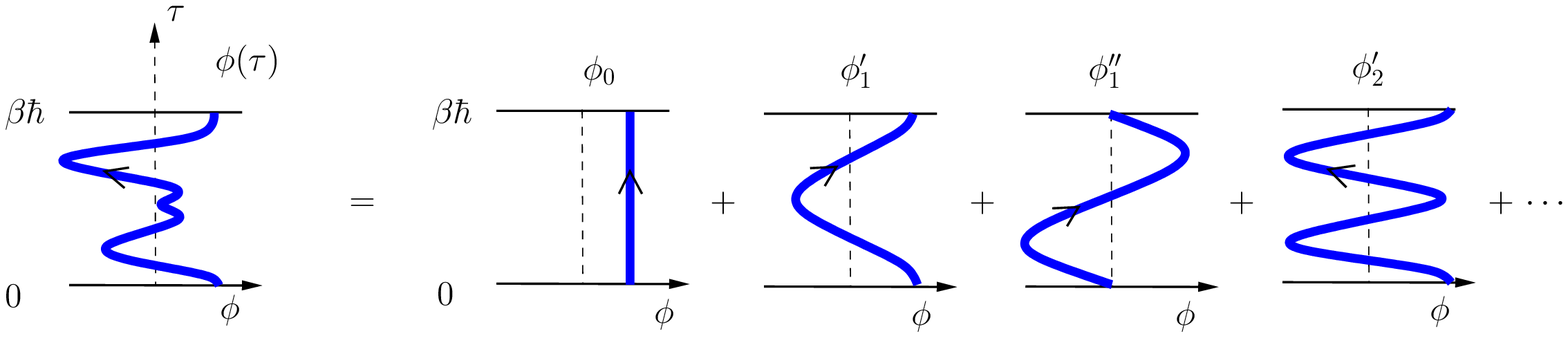}{Periodic paths in imaginary time for the harmonic
oscillator can be decomposed into their Fourier modes, with amplitudes
$\phi_{n}= \phi_{n}'+i\phi_{n}''.$}{pedfig4}

Let us now see how boundary conditions in time cause the system to
become ``hot'' for our example. More specifically 
consider the variance in the
displacement 
\begin{equation}
\langle \hat \phi^{2}\rangle =
\frac{1}{\hbar \beta }\!\int_{0}^{\hbar \beta }d\tau 
\langle \hat \phi^{2} (\tau)\rangle = \frac{1}{\hbar \beta }
\sum_{n}\langle |\phi_{n}|^{2}\rangle = 
k_B T\sum_{n} \frac{1}{m (\omega_{0}^{2}+\nu_n^2)},
\end{equation}
which has been rewritten in terms of the normal modes 
so that the average potential energy is given by
\begin{equation}
\langle \hat V\rangle = \frac{m\omega_{0}^{2}}{2}\langle
\phi^{2}\rangle 
= \frac{\omega_{0}^{2}}{2}k_B T\sum_{n} \frac{1}{ (\omega_{0}^{2}+\nu_n^2)}.
\end{equation}
We can calculate this expression
using contour integration methods. We
rewrite it as a clockwise contour integral around the poles of $n (z)=
[e^{\hbar z / k_{B}T}-1]^{-1}$ at $z= i\nu_n$ (Fig.~\ref{fig5}(a))
\begin{subequations}
\begin{align}
\langle \hat V\rangle &=\frac{\hbar \omega_{0}^{2}}{2} \oint_{C} \frac{dz}{2 \pi i }n_{B} (z)\frac{1}{(\omega_{0}^{2}-z^{2})}\\
&=\frac{\hbar \omega_{0}^{2}}{2} \oint_{C'} \frac{dz}{2 \pi i }n_{B}
(z)\frac{1}{(\omega_{0}^{2}-z^{2})
},
\end{align}
\end{subequations}
where $C'$ runs clockwise around the poles at $z=\pm \omega_{0}$
(Fig.~\ref{fig5}(b)). 
\fg{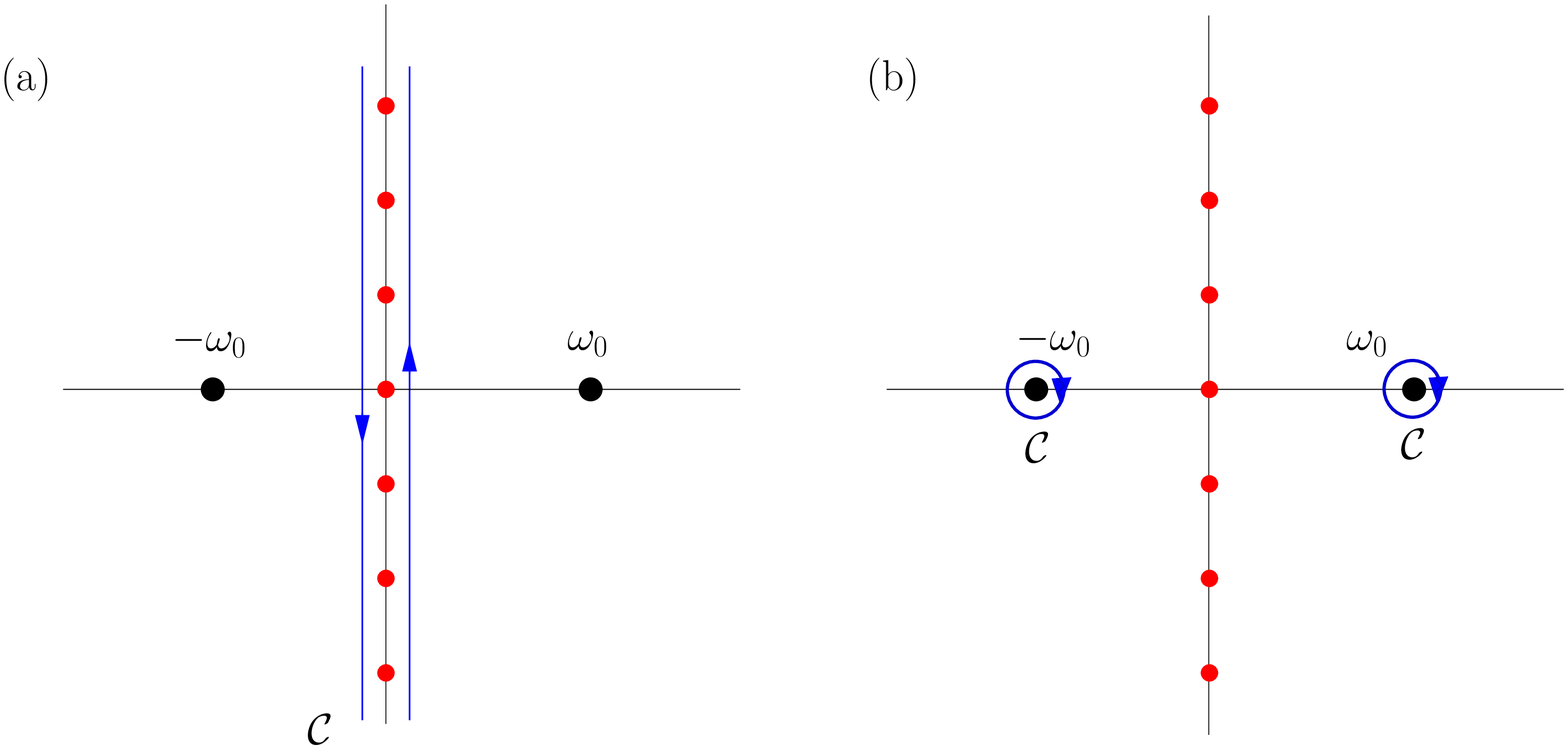}{(a) The contour integration path $\cal C$ used to
evaluate Eq.~(\ref{bbodyint}). (b) The distortion of the contour to form the path $\cal
C'$ around the poles at
$z=\pm \omega_0$.
}{fig5}
The
resulting integral is
\begin{subequations}
\label{bbodyint}
\begin{align}
\langle \hat V\rangle &=\frac{\hbar \omega_{0}^{2}}{2} \left(
\frac{n_B
(\omega_{0})}{2\omega_{0}}+ 
\frac{n_B
(-\omega_{0})}{-2\omega_{0}} \right)\cr
&=\frac{\hbar \omega_{0}}{2}\left[n_{B} (\omega_{0})+\frac{1}{2} \right].
\end{align}
\end{subequations}
From the virial theorem\cite{Pathria72} we expect
a similar expression for the kinetic energy so that the
total energy of the harmonic oscillator is 
\begin{equation}
\label{TVeq}
\langle \hat T + \hat V \rangle = 
{\hbar \omega_{0}}\left[n_{B} (\omega_{0})+\frac{1}{2} \right].
\end{equation}
The second term is due to $T=0$ zero-point fluctuations, and the
first describes the thermal excitations of the oscillator. From this derivation
we see that the latter
result from the redistribution of normal mode energies, and are a direct
consequence of 
the imposition of a boundary condition in (imaginary) time. 

Blackbody radiation is a well-known example of thermally excited oscillators. 
We can easily generalize Eq.~(\ref{TVeq})
to the electromagnetic vacuum, 
an ensemble of 
harmonic oscillators, by replacing $\omega_0\rightarrow c q$, so that 
the energy density at a finite temperature is given by
\begin{subequations}
\begin{align}
\frac{\cal E_T}{V} &= 
\frac{1}{V}\sum_{\bq} 2 \hbar c q \left[n_B(cq)+\frac{1}{2}\right]\cr
&=
2 \!\int \frac{d^3 q}{(2 \pi)^3}\hbar c q \left[n_B(cq)+\frac{1}{2}\right],
\end{align}
\end{subequations}
where the factor of two is due to the different 
polarizations. With the substitution $x= \hbar c q\beta$, the thermal 
energy density takes the form
\begin{align}\label{coefficient}
\frac{\Delta {\cal E}_T}{V} = 
\frac{(k_BT)^4}{\pi^2 c^3 \hbar^3}\!\int dx x^{3} n_B(x).
\end{align}
We compare Eq.~\eqref{coefficient} with the energy density obtained in the 
Casimir effect in Eq.~(\ref{bbody}). 
With $\beta_C^{-1} = k_B T_C = \hbar c/2a$, the Casimir energy density becomes
\begin{equation}
\frac{\Delta{\cal E}_C}{A a}= - \frac{(k_B T_C)^4}{3 \pi^{2 }c^3 \hbar^3}\!
\int\! dx\,x^{3} n_B(x). 
\end{equation}
The similarity
between the Casimir and the blackbody energy density is testament 
to their common origin as boundary-condition effects. 
The sign difference is due to the subtle distinctions between 
imaginary and real time, and the 
consequences for the kinetic energy.
As an aside, we note that the blackbody radiation pressure,
$P=\Delta {\cal E}_T/3V$, has the same
prefactor (apart from the sign) as the 
Casimir pressure.\cite{Landau80}
Traditionally we think of black body radiation as resulting
from an excitation of thermal modes.
Our calculation shows that the Casimir effect and
blackbody radiation are both 
consequences of boundary conditions and the 
{\sl redistribution} of zero-point fluctuation modes in the vacuum. 

Recent experiments have observed the Casimir effect between
parallel plates with a one micron 
separation ($a=1\,\mu$m).\cite{Lamoreaux97,Mohideen98,Chan01,Lisanti05,Obrecht07,Munday09} The corresponding
``Casimir temperature'' for these experiments is
\begin{equation}
T_{C}= \frac{\hbar c }{2 a k_{B}} \sim 1000\,\mbox{K}.
\end{equation}
The Casimir effect at these length scales couples to
the same photons that predominate in the blackbody spectrum at 1000\,K.
The boundary conditions imposed by the two phenomena on the
electromagnetic field are almost identical.

More generally, zero-point fluctuations play a major role 
at quantum phase 
transitions.\cite{Sondhi97,Sachdev99,Coleman05} 
The effect of finite temperature in the vicinity of a ($T=0$)
quantum critical point is the temporal analog of the Casimir
phenomenon, a ``Casimir effect in time,'' where temperature imposes temporal
constraints on critical zero-point fluctuations. 
As we have discussed, there is an intimate connection 
between a finite temporal dimension and a
nonzero temperature in a quantum system,
\cite{Cardy96,Sondhi97,Sachdev99,Continentino01,Coleman05} and this connection
has many observable consequences on thermodynamic
quantities for quantum critical systems 
at nonzero temperatures.\cite{Palova08}
Heuristically, this link between temperature and a boundary condition in time
can be understood within the
framework of the Heisenberg uncertainty principle
\begin{equation}
\label{uncertainty}
\Delta t \sim \frac{\hbar}{k_B T},
\end{equation}
where a thermal energy fluctuation 
leads to an upper cutoff in time, the Planck time,
that is inversely proportional to the temperature.
More formally, as we have seen in our simple example,
finite-temperature emerges in a path integral framework as a periodic boundary effect 
in imaginary time, which becomes particularly important
near a quantum critical point where there
exist quantum fluctuations on all spatial
and temporal scales. 
Here finite-temperature
corresponds to the redistribution of quantum zero-point fluctuations
due to the imposition of external constraints, and
thus is analogous to the Casimir effect for two parallel 
metallic plates in vacuum. 
Running this argument the other way, we note that
the removal of temporal modes by periodic finite boundary conditions
generates a temperature (and thus entropy and thermal energy) in a system near a quantum 
critical point.
As an aside, 
we remark that finite-temperature
effects resulting from boundary constraints
have been discussed in the context of
astrophysics where 
blackbody radiation and event
horizons have been linked via the Unruh
effect.\cite{Unruh76,Thorne95}

In conclusion, we have revisited the Casimir effect
with an approach used to study condensed matter systems
at finite temperature. We recovered results previously
derived by other methods and also discussed the physical
implications of analogies implicit in this treatment.
Our hope is that this presentation will
make this topic straightforward to include in a 
graduate many-body course for future condensed matter physicists.

\begin{acknowledgments}

This perspective on the Casimir effect emerged from research
on quantum paraelectrics, funded by the DOE (grant DE-FE02-00ER45790),
and research on finite-size phenomena in quantum systems, supported by the NSF 
(grants NSF-DMR 0645461 and NSF-NIRT-ECS-0608842).
\end{acknowledgments}


\begin{thebibliography}{99}

\bibitem{Casimir48} H. B. G. Casimir, ``On the attraction between two 
perfectly conducting plates,'' Proc. Kon. Ned. Akad. Wetenschap 
{\bf 51}, 793--795 (1948).

\bibitem{Lambrecht02} A. Lambrecht, ``The Casimir effect: A force
from nothing,'' Physics World {\bf 15} (9), 29--32 (2002).

\bibitem{Lamoreaux07} S. K. Lamoroeaux, ``Casimir forces: Still surprising
after 60 years,'' Phys. Today {\bf 60} (2), 40--45 (2007).

\bibitem{Itzykson80} C. Itzykson and J.-B. Zuber, {\sl Quantum Field Theory} 
(McGraw-Hill, New York, 1980).

\bibitem{Lamoreaux97} S. K. Lamoreaux, ``Demonstration of the Casimir force
on the 0.6 to 6\,$\mu$mm range,'' Phys. Rev. Lett. {\bf 78}, 5--8
(1997).

\bibitem{Mohideen98} U. Mohideen and A. Roy, ``Precision measurement of the Casimir force from 0.1 to 0.9\,$\mu$m,'' 
Phys. Rev. Lett. {\bf 81}, 4549--4552
(1998).

\bibitem{Chan01} H. B. Chan, V. A. Aksyuk, R. N. Kleiman, D. J. Bishop, and F. Capasso, ``Quantum mechanical actuation of micromechanical systems by the Casimir force,'' Science {\bf 291}, 1941--1944
(2001).

\bibitem{Lisanti05} M. Lisanti, D. Iannuzzi, and F. Capasso, ``Observation of the skin-depth effect on the Casimir force between metallic surfaces,'' PNAS {\bf 102}, 11989--11992
(2005).


\bibitem{Obrecht07} J. M. Obrecht, R. J. Wild, M. Antezza, L. P. Pitaevskii, 
S. Stringari, and E. A. Cornell, ``Measurement of the 
temperature-dependence of the Casimir-Polder force,'' Phys. Rev. Lett. {\bf 98}, 063201-1--4 (2007).


\bibitem{Munday09} J. N. Munday, F. Capasso, and V. A. Parsegian, ``Measured long-range repulsive Casimir-Lifshitz forces,'' {\sl Nature} {\bf 457}, 170--173 (2009).

\bibitem{Fisher78} M. E. Fisher and P. G. deGennes, C.R. Acad. Sci. Paris B {\bf 287}, 207--209 (1978).

\bibitem{Krech94} M. Krech, {\sl The Casimir Effect in Critical Systems}
(World Scientific, Singapore, 1994).

\bibitem{Kardar99} M. Kardar and R. Golestanian, ``The `friction' 
of vacuum, and other fluctation-induced forces,'' 
Rev. Mod. Phys. {\bf 71}, 1233--1245 (1999).

\bibitem{Danchev00} D. M. Danchev, J. G. Brankov and N. S. Tonchev, {\sl Theory of Critical Phenomena in Finite-Size Systems: Scaling 
and Quantum Effects} (World Scientific, Singapore, 2000).

\bibitem{Gambassi08} A. Gambassi, ``The Casimir effect: From quantum to
critical fluctuations,'' 
J. Phys.: Conf. Ser. {\bf 161} 012037-1--17 (2009).  

\bibitem{Palova08} L. P\'alov\'a, P. Chandra, and P. Coleman, 
``Quantum critical paraelectrics and the Casimir effect in time,''
Phys. Rev. B {\bf 79}, 075101-1--17 (2009).

\bibitem{Belifante87} F. J. Belinfante, ``The Casimir effect revisited,'' 
Am. J. Phys. {\bf 55}, 134--138 (1987).

\bibitem{Dowling89} J. P. Dowling, ``The mathematics of the Casimir 
effect,'' {\sl Mathematics Magazine} {\bf 62}, 324--331 (1989).

\bibitem{Elizalde91} E. Elizalde and A. Romeo, ``Essentials of the Casimir 
effect and its computation,'' Am. J. Phys. {\bf 59}, 711--719 (1991).

\bibitem{Revzen97} M. Revzen, R. Opher, M. Opher, and A. Mann, ``Kirchoff's theorem and
the Casimir effect,'' Europhys. Lett. {\bf 38}, 245--248 (1997).

\bibitem{Lamoreaux99} S. K. Lamoreaux, ``Resource Letter CF-1: Casimir 
force,'' Am. J. Phys. {\bf 67}, 850--861 (1999).

\bibitem{Mahan07} G. D. Mahan, {\sl Many-Particle Physics} 
(Springer, New York, 2007), 3rd ed. 

\bibitem{Jackson98} J. D. Jackson, {\sl Classical Electrodynamics} (John Wiley \& Sons, New York, 1998),
3rd ed., pp. 352--363.

\bibitem{Boyer74} T. H. Boyer, ``Van der Waals forces and zero-point entropy for dielectric and permeable materials,'' Phys. Rev. A {\bf 9}, 2078--2084 (1974).

\bibitem{Hushwater97} V. Hushwater, ``Repulsive Casimir force as a result of
vacuum radiation pressure,''Am. J. Phys. {\bf 65}, 381--384 (1997).

\bibitem{Feynman65} R. P. Feynman and A. R. Hibbs, {\sl Quantum Mechanics and Path Integrals} (McGraw-Hill, New York, 1965); R. P. Feynman, ``Space-time approach to non-relativistic quantum mechanics,'' Rev. Mod. Phys. {\bf 20}, 367--387 (1948).

\bibitem{Pathria72} R. K. Pathria, {\sl Statistical Mechanics} 
(Butterworth-Heinemann, Oxford, 1996), 2nd ed.

\bibitem{Landau80} L. D. Landau and E. M. Lifshitz, 
{\sl Statistical Physics} (Pergamon Press, Oxford, 1980).

\bibitem{Cardy96} J. Cardy, {\sl Scaling and Renormalization in 
Statistical Physics} (Cambridge University Press, Cambridge, 1996).

\bibitem{Sondhi97} S. L. Sondhi, S. M. Girvan, J. P. Carini, and D. Shahar, 
``Continuous quantum phase transitions,'' Rev. Mod. Phys. 
{\bf 69}, 315--333 (1997).

\bibitem{Sachdev99} S. Sachdev, {\sl Quantum Phase Transitions} (Cambridge University Press, Cambridge, 1999).

\bibitem{Coleman05} P. Coleman and A. J. Schofield, ``Quantum criticality,'' Nature {\bf 433}, 226--229 (2005).


\bibitem{Continentino01} M. A. Continentino, {\sl Quantum Scaling in Many-Body Systems} (World Scientific, Singapore, 2001). 

\bibitem{Unruh76} W. G. Unruh, ``Notes on black hole evaporation,'' Phys. Rev. D {\bf 14}, 870--892 (1976).

\bibitem{Thorne95} K. Thorne, {\sl Black Holes and Time Warps: Einstein's Outgrageous Legacy} 
(W. W. Norton, New York, 1995).

\end{thebibliography}
\end{document}